\documentclass{aastex}
\usepackage{emulateapj5}
\usepackage{onecolfloat5}
\newcommand{\gsim}{\;\rlap{\lower 2.5pt
 \hbox{$\sim$}}\raise 1.5pt\hbox{$>$}\;}
\newcommand{\lsim}{\;\rlap{\lower 2.5pt
   \hbox{$\sim$}}\raise 1.5pt\hbox{$<$}\;}
\newcommand{\be}{\begin{equation}}
\newcommand{\beq}{\begin{equation}}
\newcommand{\ba}{\begin{eqnarray}}
\newcommand{\ee}{\end{equation}}
\newcommand{\eeq}{\end{equation}}
\newcommand{\ea}{\end{eqnarray}}

\def\vecu{{\bf u}}

\newcommand{\bea}{\begin{eqnarray}}
\newcommand{\eea}{\end{eqnarray}}
\newcommand{\bean}{\begin{eqnarray*}}
\newcommand{\eean}{\end{eqnarray*}}

\newcommand{\br}{{\bf r}}
\newcommand{\bk}{{\bf k}}
\newcommand{\bn}{{\bf \hat{n}}}

\newcommand{\nn}{\nonumber\\}

\begin{document}

\twocolumn[
\title{Multifrequency Analysis of 21 cm fluctuations From the Era of Reionization}
\author{M\'ario G. Santos}
\affil{Department of Physics, University of California, Davis.\\ 
CENTRA, Instituto Superior T\'ecnico, Universidade T\'ecnica de Lisboa, Portugal.\\
Email: santos@physics.ucdavis.edu}
\author{Asantha Cooray}
\affil{Theoretical Astrophysics, MS 130-33, California Institute of Technology, Pasadena, CA 91125.\\
Department of Physics and Astronomy, University of California, Irvine,
CA 92697.\\ 
E-mail: asante@caltech.edu}
\author{Lloyd Knox}
\affil{Department of Physics, University of California, Davis.\\ Email: lknox@ucdavis.edu}

\begin{abstract}
We study the prospects for extracting detailed statistical properties
of the neutral Hydrogen distribution during the era of reionization
using the brightness temperature fluctuations from redshifted 21 cm
line emission.  Detection of this signal is complicated by
contamination from foreground sources such as diffuse Galactic
synchrotron and free-free emission at low radio frequencies,
extragalactic free-free emission from ionized regions and radio point
sources.  We model these foregrounds to determine the extent to which
21 cm fluctuations can be detected with upcoming experiments.  We find
that not only the level of correlation from one frequency to another,
but also the functional form of the foreground correlations has a
substantial impact on foreground removal.  We calculate how well the
angular power spectra of the 21cm fluctuations can be determined.  We
also show that the large-scale bias of the neutral Hydrogen gas
distribution with respect to the density field, can be determined with
high precision and used to distinguish between different reionization
histories.
\end{abstract}

]

\section{Introduction}

Recently there has been a great deal of interest in the reionization
of the intergalactic medium
\citep{cen03,chen03,cen02,gnedin04b,wyithe03,haiman03,sokasian03,miralda-escude00}.
This interest was sparked by the detection \citep{bennett03,kogut03} of a correlation
between polarization and temperature on large angular scales \citep{zaldarriaga97a}.  
The amplitude of the signal suggests that reionization began much earlier than
$z \sim 6$, which is when the inter-galactic medium became completely
ionized as we know from the detection of Gunn-Peterson troughs \citep{gunn65,becker01,fan01}.

Now that the ionization history appears to be quite rich, we wish to
study the epoch in detail. Unfortunately, the CMB is mostly sensitive
only to the integral history of reionization.  Although different
reionization histories with the same optical depth do give different
large angular scale polarization power spectra, one cannot easily use 
these changes to fully reconstruct the reionization history as a function 
of redshift due to the large cosmic variance on large angular scales 
\citep{Kaplinghat03a,hu03,haiman03}.

Another way to study the reionization history of the Universe in more detail is through 
the observation of the distribution of neutral Hydrogen based on the spin-flip transition at a rest wavelength of 21 cm.
Plans for upcoming low-frequency radio experiments such as PAST\footnote{http://web.phys.cmu.edu/${}_{\tilde{}}$past/} \citep{pen04}, Mileura Widefield Array (MWA)\footnote{http://web.haystack.mit.edu/arrays/MWA/site}, Low Frequency Array (LOFAR)\footnote{http://www.lofar.org} and Square Kilometer Array (SKA)\footnote{http://www.skatelescope.org}
have now motivated detailed study of the 21 cm background associated with neutral hydrogen during, and prior
to, reionization (e.g., \citealt{Scott90,madau97,gnedin97,zaldarriaga04,morales03,kumar95,tozzi00,iliev02,Bharadwaj04}). 
Unlike the CMB, the advantage here is that by a priori selecting the observed frequency, 
one can probe the neutral content of the Universe at a given redshift, thereby obtaining a
tomographic view of the reionization process. This way, one can
reconstruct the reionization history and determine, especially, if the reionization 
process was either instantaneous or lengthy and complex
as implied by current CMB data. Moreover, since observable effects in CMB data depend only
on the ionized content while  21 cm fluctuations come from inhomogeneities in the neutral distribution, the combination
may allow additional information in the study of reionization.

Although there is strong scientific motivation to study 21 cm fluctuations during the era of reionization, there are
several difficulties to overcome. In addition to experimental challenges, involving observations at low radio frequencies,
the background itself is highly contaminated by foreground radio emission.
Synchrotron and free-free emission from the Milky Way \citep{shaver99}, low frequency radio point sources
\citep{dimatteo02,dimatteo04} and free-free emission from free electrons in
the intergalactic medium \citep{oh99,cooray04} are now thought to be
the chief sources of confusion. While it has been suggested \citep{oh03} that background studies may be impossible due to
the large number of contaminating sources (with brightness fluctuations much higher than expected for neutral Hydrogen
at high redshifts), others have contended that
a multifrequency analysis of the 21 cm data can be used to reduce the foreground contamination given the  smoothness of
these contaminants in frequency space \citep{shaver99,dimatteo02,zaldarriaga04}. Initial calculations suggest that, for 
experiments such as LOFAR and SKA, one can easily 
remove the foregrounds to a sufficiently low level that
statistical studies related to 21 cm fluctuations can be achieved \citep{zaldarriaga04}.

In this paper, we extend the initial discussions related to foreground removal by considering a detailed analysis of the
multifrequency removal technique. 
When estimating how accurately cosmological parameters can be measured, we take into account the complication
that most foreground properties, like the level of smoothness across frequency, must be estimated at the same 
time as the signal. Our foreground model includes not only the normalization, frequency dependence and 
scale dependence for each physical quantity, but also variations in the frequency coherence and even 
the functional form of the foreground correlations across frequency.
We also consider the extent to which 21 cm
fluctuations can be cross-correlated between frequency bins. This cross-correlation of the signal between channels
reduces the efficiency with which foregrounds can be cleaned.

The paper is organized as follows: in \S \ref{21cm}, we discuss
the 21 cm signal anisotropy and frequency correlations.
In \S \ref{foreg}, we give an overview of the most important foreground contaminants.
In \S \ref{exp_setup} we discuss the experimental setup assumed in our analysis. In \S \ref{forecast}
we discuss the method used for the error forecast and
the multifrequency cleaning techniques to reduce the foreground
contamination. Finally in \S \ref{reion} we 
study how well physical parameters related to the 21 cm signal can be extracted.
Throughout the paper, we make use of the WMAP-favored $\Lambda$CDM cosmological model \citep{spergel03}.

\section{The 21cm signal}
\label{21cm}

When traveling through a patch of neutral hydrogen, the intensity of the CMB radiation will
change due to absorption and emission. The corresponding change in the brightness temperature,
$T_{21}(\nu)$, as compared to the CMB at an observed frequency $\nu$ is then
\bea
T_{21}(\nu) & \approx & \frac{T_S - T_{\rm CMB}}{1+z} \, \tau 
\label{eq:dtb} 
\eea
where $T_S$ is the temperature of the source (the spin temperature of the IGM), $z$ is the
redshift corresponding to the frequency of observation ($1+z=\nu_{21}/\nu$, with 
$\nu_{21} = 1420$ MHz) 
and $T_{\rm CMB} = 2.73 (1+z) K$ is the CMB temperature at redshift $z$.
The optical depth, $\tau$, of this patch in the hyperfine transition \citep{field59} is given 
in the limit of $k_B T_s >> h \nu_{21}$ by
\bea
\tau & = & \frac{ 3 c^3 \hbar A_{10} \, n_{\rm HI}}{16 
k \nu_{21}^2 \, T_S \, H(z) } 
\label{eq:tauigm} \\
\, & \approx & 8.6 \times 10^{-3} (1+\delta_b) x_H \left[
\frac{T_{\rm CMB}(z)}{T_S} \right] 
\left( \frac{\Omega_b h^2}{0.02} \right) \nonumber \\
\, & \, & \times
\left[ \left(\frac{0.15}{\Omega_m h^2} \right) \, \left(
\frac{1+z}{10} \right) \right]^{1/2} \left( \frac{h}{0.7}
\right)^{-1}, 
\nonumber
\eea
where $A_{10}$ is the spontaneous emission coefficient for the transition ($2.85 \times 10^{-15}$ s$^{-1}$)
and $n_{\rm HI}$ is the neutral hydrogen density. This can be expressed as
$n_{\rm HI}=x_H\bar{n}_b(1+\delta_b)$, when $\bar{n}_b$ is the mean number density of cosmic baryons, with a
spatially varying overdensity $\delta_b$ and
$x_H$ is the fraction of neutral hydrogen ($x_H= 1-x_e$ where $x_e$ is the fraction of
free electrons).
We refer the reader to \citet{zaldarriaga04} for further details.

\subsection{21cm Power Spectrum}
                                 
The measured brightness temperature corresponds to a convolution of the intrinsic
brightness with some response function $W_\nu(r)$ that characterizes the bandwidth of the
experiment:
\be
T^s(\bn,\nu_0)=\int dr W_{\nu_0}(r) T_{21}(\bn,r),
\ee
where $\bn$ is the direction of observation and $r$ is the radial distance corresponding
to the observed frequency $\nu$.                                                                                            
We can now determine the angular power spectrum of $T^s(\bn,\nu_0)$ which is related to the 3-d
power spectrum of $T_{21}(\bn,r)$, defined by
\be
\label{3dpower}
\langle\tilde{T}_{21}(\bk,\nu_1)\tilde{T}_{21}(\bk',\nu_2)\rangle=
(2\pi)^3\delta^D(\bk+\bk') P_{21}(k,\nu_1,\nu_2),
\ee
where $T_{21}(\bn,r)\equiv T_{21}(\br,\nu)=
\int {d^3k\over (2\pi)^3} \tilde{T}_{21}(\bk,\nu) e^{i\bk\cdot\br}$.
Making use of the spherical harmonic moment of the 21 cm fluctuations, at a frequency $\nu_0$,
\begin{equation}
a^s_{l m}(\nu_0) = \int d\bn Y^*_{l m}(\bn) T^s(\bn,\nu_0) \, ,
\end{equation}
we write the angular power spectrum as (e.g. \citealt{kaiser92})
\bea
\label{cl21}
& &\langle a^s_{l m}(\nu_1) a^{s*}_{l m}(\nu_2)\rangle=C^s_l(\nu_1,\nu_2) \\
& &={2\over\pi}\int k^2 dk\,P_{21}(k,\nu_1,\nu_2) I_l^{\nu_1}(k) I_l^{\nu_2}(k) \, , 
\nonumber
\eea
where 
\begin{equation}
I_l^{\nu}(k) = \int dr W_{\nu}(r) j_l(k r) \, ,
\end{equation}
with the spherical Bessel function given by $j_l(k r)$. In deriving this form for the angular power
spectrum, we have made use of the Rayleigh expansion for the plane wave given by
\begin{equation}
e^{i\bk\cdot\br}= 4\pi \sum_{l m} i^{l} j_l(k r) Y_l^{m}(\bn) Y_l^{m*}(\hat{\bk}) \, .
\end{equation}

\begin{figure}[t]
\includegraphics[scale=0.4,angle=-90]{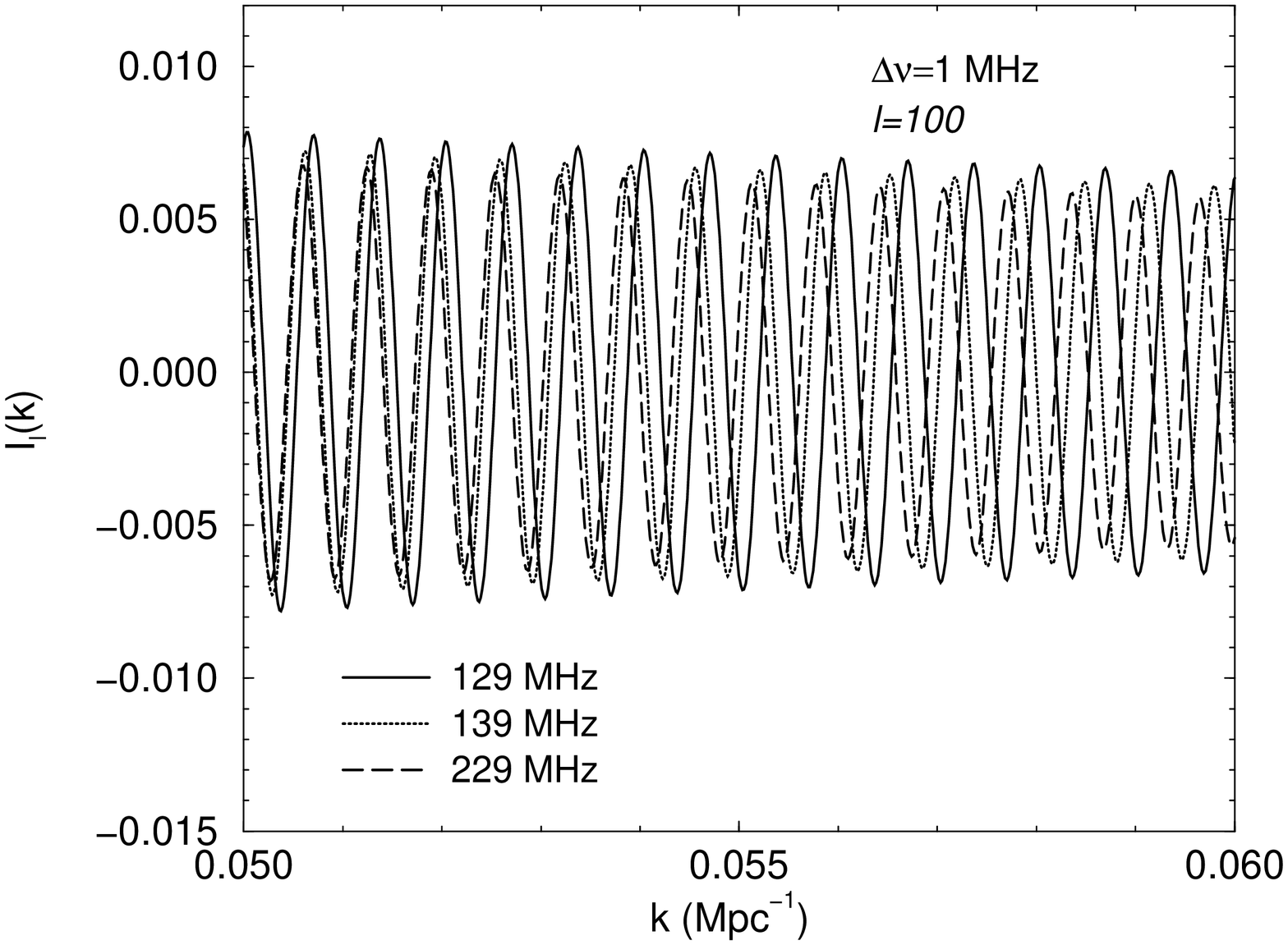}
\includegraphics[scale=0.43,angle=-90]{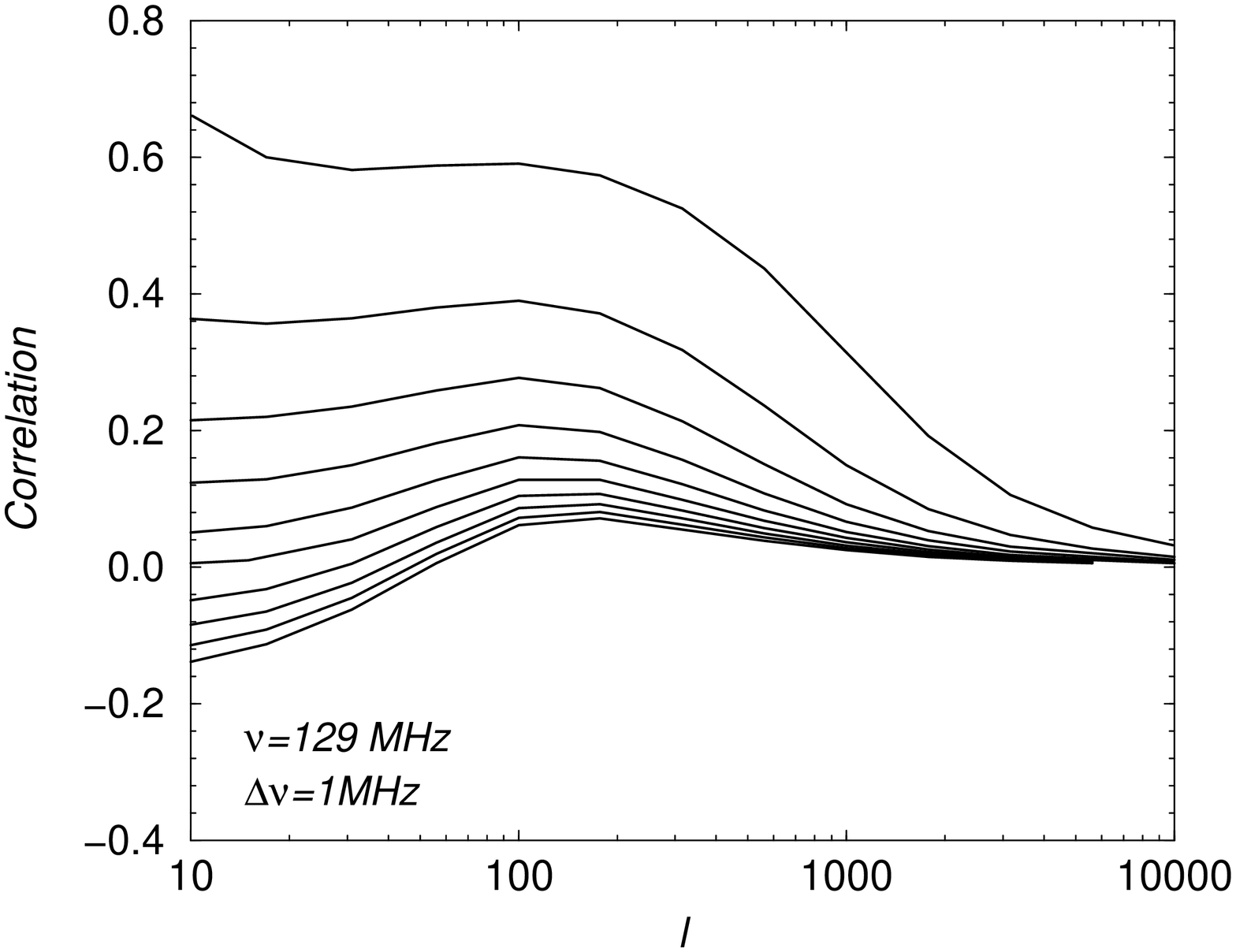}
\caption{The frequency cross-correlations of the 21cm signal.
Top panel: the absolute value of the 
function $I_l^\nu(k)$ when $l=100$ for three frequency bins with a width of 1MHz.
For large frequency separations these functions
are out of phase and the integral over k is essentially zero. Bottom panel: signal correlations across frequency. Each curve
corresponds to a frequency separation from 1MHz to 10MHz (1MHz bandwidth).}
\label{fig:corr2}
\end{figure}

In Figure \ref{fig:corr2}, we show in the top panel the integral, $I_l^\nu(k)$, when $l=100$ 
as a function of $k$, for three different
frequency bins of width 1MHz.
As the frequency separation increases, 
the oscillations are out of phase suggesting that $C^s_l(\nu_1,\nu_2)$ is not significant for $\nu_2-\nu_1 \gg 10{\rm MHz}$.
However, for smaller separations $C^s_l(\nu_1,\nu_2)$ certainly cannot be 
considered to be zero. This is clear by looking at the bottom panel of Figure \ref{fig:corr2} where we show
the signal correlations across frequency for separations from 1 to 10MHz. 

To proceed further we need an expression for the power spectrum of the 21 cm fluctuations, $P_{21}(k,z)$,
or in this case, spatial inhomogeneities in the underlying gas density field.
In order to simplify the calculation related to 
fluctuations in the density field, we assume 
a scenario where the gas density field is heated well above the CMB temperature
during the reionization process.
In addition to reionization, we also expect the structure formation process,
such as shocks in forming virialized dark matter halos, where first stars eventually form to subsequently reionize the universe,
to heat the gas such that $T_s\gg T_{CMB}$ \citep{gnedin04}. The gravitational shocks will also heat the gas in overdense regions like
sheets and filaments such that the assumption we make here with regards to the gas temperature should be valid throughout 
the IGM at a time around the era of reionization where one also expects a substantial background of X-rays from
supernovae, etc \citep{venkatesan01,chen04} to also heat the IGM quickly.
At sufficiently high $z$, prior to significant reionization and heating, we expect $T_s < T_{\rm CMB}$
and it may be possible to see HI in 21cm absorption \citep{madau97}.

With $T_s >> T_{\rm CMB}$, the 21 cm signal will be observed in emission with respect to the background CMB.
Writing the brightness temperature as:
\be
T_{21}(z)=c\,(1+\delta)(1-\bar{x}_e-\bar{x}_e\delta_x),
\ee
where 
\be
c\approx 23\left( \frac{\Omega_b h^2}{0.02} \right) \nonumber \\
\left[ \left(\frac{0.15}{\Omega_m h^2} \right) \, \left(
\frac{1+z}{10} \right) \right]^{1/2} \left( \frac{h}{0.7}
\right)^{-1} {\rm mK}
\ee
and $\delta_x$ is the perturbation in the ionization fraction ($\delta_x\equiv {x_e-\bar{x}_e\over\bar{x}_e}$),
the corresponding three-dimensional power spectrum is
\bea
\label{p21}
P_{21}(k,z)&=&c^2\left[(1-\bar{x}_e)^2 P_{\delta \delta}(k,z)+\bar{x}_e^2 P_{\delta_x \delta_x}(k,z)\right.\nn
&-& \left. 2 P_{\delta \delta_x}(k,z)\bar{x}_e(1-\bar{x}_e)\right].
\eea
The dark matter power spectrum is represented by $P_{\delta \delta}(k,z)$ (and we are assuming that 
$\delta=\delta_b$), $P_{\delta_x \delta_x}(k,z)$
is the power spectrum from the perturbations in the ionized fraction
and $P_{\delta \delta_x}(k,z)$ is the cross-correlation power.

The correlations in the ionized fraction depend on the reionization history. To proceed further we
used the model of patchy reionization in \citet{santos03b} to write
\be                                                                                                        
\label{patchy}
P_{\delta_x \delta_x}(k,z)=b^2(z)P_{\delta \delta}(k,z) e^{-k^2 R^2}
\ee
where $b(z)$ is a mean bias (halo bias weighted by the different halo properties) and $R$ the mean radius of the
HII patches.

We crudely model the time-dependence of $R$ as $R=\left({1\over 1-\bar{x}_e}\right)^{1/3} R_p$ 
where $R_p$ is the (comoving) size of the fundamental patch which we take to
be $R_p\sim 100$ Kpc.  The cutoff scale $R$ is equal to $R_p$ when reionization starts and
increases with time as HII regions overlap and form larger HII regions.  The dependence of
$R/R_p$ on $\bar x_e$ follows from assuming a Poisson distribution of patches and that their
volumes add when they overlap.  The essential consequence of this dependence of $R$ on $x_e$
and our choice of $R_p$ is that for $l \la 10,000$ (scales greater than 1 Mpc) 
the transition is very sudden; i.e., $R$ increases from 1 Mpc to $\infty$ very rapidly.

In reality there is a distribution of patch sizes and there
are spatial correlations between ionizing sources which cause some of the patches to overlap
earlier than they would given a Poisson distribution. \citet{furlanetto04a} have recently 
developed a sophisticated model of the
size distribution of patches and their correlations. 
They find a much more gradual transition to large patch sizes.  In their model, large patch sizes suppress
fluctuation power even as low as $l=1000$ while the ionization fraction is still as low as 
$\bar x_e = 0.5$ (although this depends on the assumed efficiency with which the collapsed baryons ionize the
surrounding hydrogen). 
We caution the reader that modeling the signal
on scales smaller than the largest patches is difficult and
the shape of the signal on these scales remains highly uncertain\footnote{Note also that for a 1MHz bandwidth
the signal will be smoothed out on scales smaller than $\sim 18$Mpc at $z=10$.}.

\begin{figure}[!bt]
\vspace{-0.5cm}
\plotone{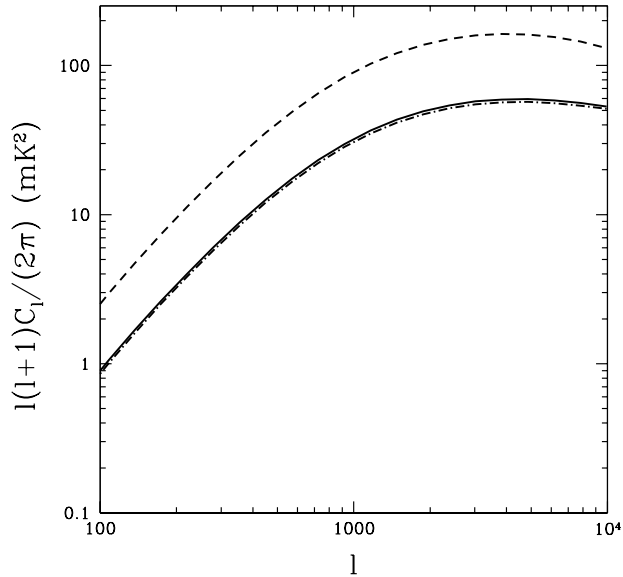}
\caption{The power spectrum for the 21cm signal at z=9.2 and assuming a bandwidth of
$\Delta \nu=1$MHz.
The solid, dashed and dot-dashed lines use the corresponding reionization models 
shown in Figures~\protect\ref{fig:xe} and \protect\ref{fig:bias}.} 
\label{fig:power21}
\end{figure}

\begin{figure}[!bt]
\vspace{-0.2cm}
\plotone{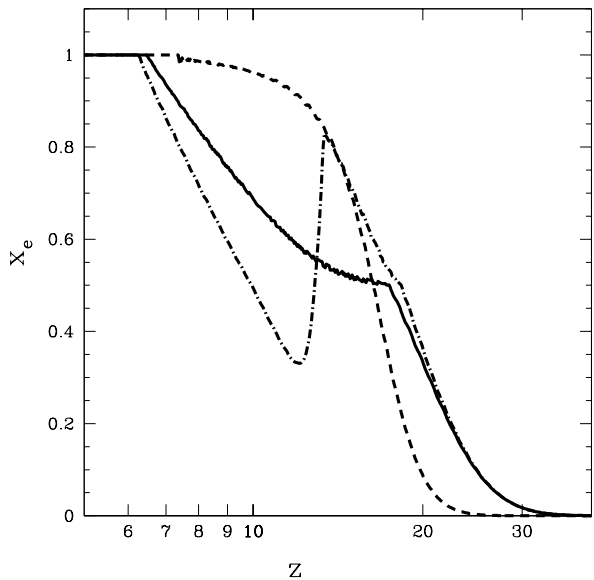}
\caption{$\bar{X}_e$ for three different fiducial models with $\tau=0.17$.
Solid line is for the fiducial model used throughout the paper. Dashed line assumes
that $H_2$ feedback effectively shuts off star formation in all Type II halos.
Dot-dashed line assumes that a transition from metal free to normal star formation takes
place abruptly at a certain redshift.}
\label{fig:xe}
\end{figure}

\begin{figure}[!bt]
\vspace{-0.2cm}
\plotone{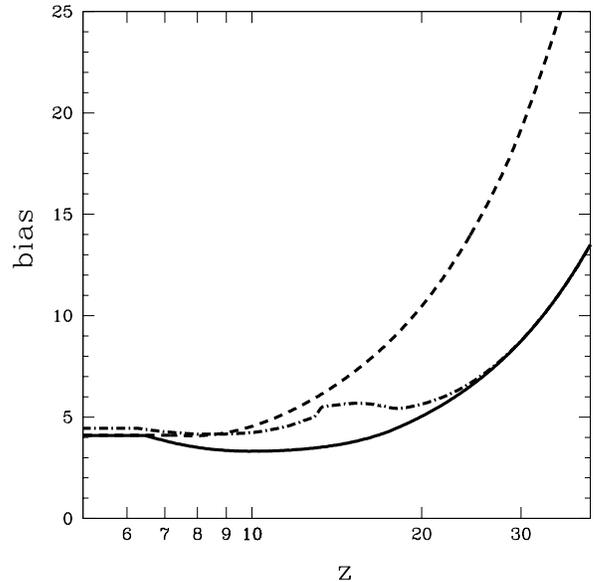}
\caption{Effective bias
for the same fiducial models as in Figure~\protect\ref{fig:xe}.} 
\label{fig:bias} 
\end{figure}

The bias in equation \ref{patchy} is much larger than 1 so that we can safely neglect
the cross-correlation contribution in our analysis.
We show in Figure \ref{fig:power21}
the 21cm power spectra for three reionization models with the same optical depth ($\tau=0.17$). 
Typically, for $z>7$ the reionization models give
$R\sim 0.2$Mpc so that the size of the patches are only important for $l>10^4$ and below that, the
power spectrum just follows the shape of the matter density perturbations.
Note that the bend in the curves at $l\sim r/\delta r\sim 500$, where $\delta r$ is the width of the window 
function, $W_\nu(r)$,
is due to the Limber regime where perturbations are smoothed out.
Figures \ref{fig:xe} and \ref{fig:bias} show the ionization fraction and bias for the corresponding reionization
models from \citet{haiman03}.


\section{Foregrounds}
\label{foreg}

We consider four different foregrounds: Galactic synchrotron, Galactic free-free,
extra-galactic diffuse free-free and extra-galactic point sources.
All of these contaminants produce much more fluctuation power at each
frequency than the 21cm signal.  
Measurement of the 21cm signal would be impossible if not for
the high coherence of the
contaminants across frequency, compared to the very short frequency
space correlation length of the signal.

With $i$ and $j$ labeling the different contaminants let us define
\be
C_l^{ij}(\nu_1,\nu_2) \equiv \langle a_{lm}^i(\nu_1) a_{lm}^j(\nu_2) \rangle.
\ee
We expect correlations between the different contaminants to be
negligible.
We model the foregrounds as power laws in both $l$ and $\nu$ so that:
\be
\label{eqn:fg1}
C_l^{ii}(\nu,\nu) = A_i (1000/l)^{\beta_i}(\nu_f/\nu)^{2 \bar
\alpha_i}.
\ee
where $\nu_f = 130 {\rm MHz}$.  
We also expect the foregrounds to be highly coherent,
i.e. $I_l^{ii}(\nu_1,\nu_2) \simeq 1$ where 
\be
I_l^{ii}(\nu_1,\nu_2) \equiv
C_l^{ii}(\nu_1,\nu_2)/\sqrt{C_l^{ii}(\nu_1,\nu_1)C_l^{ii}(\nu_2,\nu_2)}.
\label{corr_def}
\ee
Although we consider other parameterizations of $I_l^{ii}$, our starting
point is
\be
\label{gauss_corr}
 I_l^{ii}(\nu_1,\nu_2) = exp\left[-\log^2(\nu_1/\nu_2)/2/\xi_{ii}^2\right],
\ee
which, for the frequency range and values of $\xi_{ii}$ we will be considering,
reduces to
\be
\label{fct_corr}
I_l^{ii}(\nu_1,\nu_2) \sim 1-\log^2(\nu_1/\nu_2)/2/\xi_{ii}^2.
\ee
This form describes the departure from complete correlation to lowest 
non-trivial order.  It follows if one assumes the underlying 
sources have power-law
spectra with varying spectral indices $\alpha$.  If sources
with different spectral indices have $C_l$'s with different shapes (for
example, because they have different redshift distributions), then
$I_l$ can be $l$-dependent. Note that, due to this decorrelation,
the frequency dependence of the foreground brightness temperature will actually 
have departures from a power law and change with position on the sky.

If $C_l$ is dominated by Poisson fluctuations,
then $\xi = 1/\Delta \alpha$ where $\Delta \alpha = \langle
(\alpha - \bar \alpha)^2\rangle^{1/2}$.  This was the case considered
by \citet{tegmark98}.  If $C_l$ is dominated by clustering and $d\ln
C_l/d\ln \alpha=0$ then $\xi = \sigma_\alpha/\Delta \alpha^2$ where,
following \citet{zaldarriaga04}, we assume that correlations between
sources with $\alpha_1$ and $\alpha_2$ fall off as
$\exp{[-(\alpha_1-\alpha_2)^2/ (2\sigma_\alpha^2)]}$.

\begin{figure}
\begin{center}
\plotone{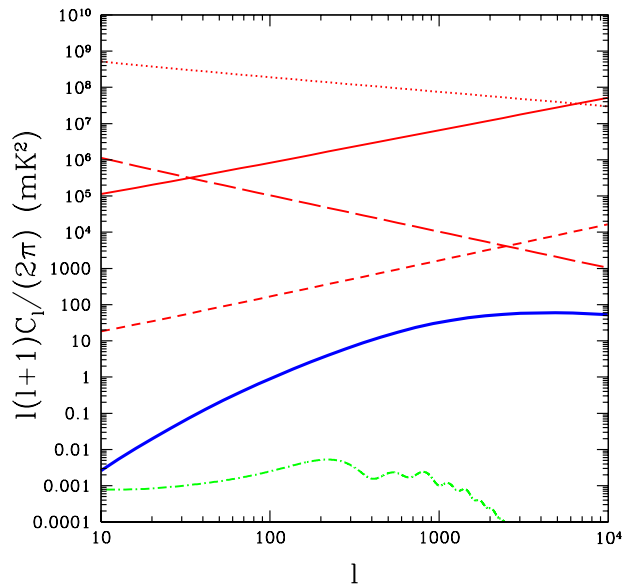}
\caption{The 21cm power spectrum (blue thick line) at z=9.2 ($\nu=140$ MHz) and its foregrounds. 
Point sources - red thin solid line, Galactic Synchrotron - red thin dotted line, 
Extragalactic free-free - red thin dashed line and Galactic free-free - red thin long-dashed line. 
CMB - green dot-dashed line.}
\label{fig:foreg}
\end{center}
\end{figure}

In Table~\ref{tab:fiducial} we show the parameters of our fiducial
models for the four different contaminants that we now discuss.
Figure~\ref{fig:foreg} shows the expected foreground power spectrum
together with the signal at 140MHz.

\begin{table}
\begin{center}
\begin{tabular}{l|c|c|c|c}
\hline 
&$A ({\rm mK}^2)$ & $\beta$ & $\bar \alpha$  & $\xi$ \\
\hline 
extragalactic point sources   &  57.0  & 1.1 &  2.07 & 1.0 \\ 
extragalactic free-free        &  0.014  & 1.0 &  2.10 & 35 \\ 
Galactic synchrotron          &  700  & 2.4 &  2.80 & 4.0 \\ 
Galactic free-free          &  0.088  & 3.0 &  2.15 & 35 \\ 
\hline
\end{tabular}
\end{center}
\caption{Fiducial parameters at $l=1000$ and $130$MHz. Parameters are defined in equations \ref{eqn:fg1}, ~\ref{corr_def}
and \ref{gauss_corr}. }
\label{tab:fiducial}
\end{table}

\subsection{Extragalactic point sources}

The extragalactic sources of contamination likely to be important at
the relevant low flux levels are radio galaxies, active galactic
nuclei and normal galaxies.  There are several surveys of extragalactic
sources in or near the wavelength range of interest including the 6C survey
\citep{hales98} at 151 MHz, the 3CR survey at 178 MHz \citep{laing83} and
a survey by \citet{cohen04} at 74 MHz.  

All but the \citet{cohen04} survey were used by \citet{dimatteo02} to
estimate the angular power spectra from unresolved sources at 150 MHz.
They model both the Poisson contribution and the correlated
contribution, assuming the sources are clustered like Lyman break
galaxies \citep{giavalisco98} at $z \sim 3$.  They find that if sources
brighter than $S_{\rm cut} = 0.1$ Jy can be detected and removed, then
the clustering signal, and not the Poisson signal, will be the
dominant contribution to the variance of maps smoothed on angular
scales larger than about 1'.  As the flux cut is reduced further, the
Poisson contribution drops faster than the clustering contribution, due
to the Poisson contribution's heavier weighting of the brightest
sources.

We use the \citet{dimatteo02} model with $S_{\rm cut}=0.1$ mJy for
the angular power spectrum of the background of unresolved sources 
\footnote{Note that the model for the differential source counts is 
$dN/dS=4(S/1Jy)^{-1.75} Sr^{-1} mJy^{-1}$}.
This translates into $A = 57{\rm mK}^2$, and $\beta = 1.1$ as shown in
Table~\ref{tab:fiducial}. 

For the frequency dependence we turn to \citet{cohen04} who
compare fluxes of 947 of their 949 sources that appear also in 
the NVSS survey at 1400 MHz \citep{condon98}.  Assuming the spectrum of
each source follows a power-law in intensity, $I_\nu \propto \nu^\gamma$, 
they find an average power-law index of
$\bar \gamma^{1400}_{74}=-0.79$ with a scatter of about 
$\delta \gamma = 0.2$.  They also see a trend for flattening of spectral
index as the 74 MHz flux density drops from 1 Jy down to 0.2 Jy going
roughly as 
\be
\bar \gamma^{1400}_{74} = -0.83 -0.13 \log_{10}\left(I_{74}/{\rm Jy}\right).
\ee
Extrapolating further down 
to 0.1 mJy we find $\bar \gamma^{1400}_{74}=-0.31$.  

\citet{cohen04} also find 545 of the 947 sources in the WENSS catalog
at 325 MHz \citep{rengelink97}.  They find a significant flattening of
the spectra of most sources towards lower frequencies.  The mean
spectral indices are related by $\bar \gamma^{325}_{74}=0.24 + \bar
\gamma^{1400}_{74}=-0.07$.  

Since fluctuations in antenna temperature follow as $\delta T (\nu) \propto \nu^{-2} \delta I_\nu$, 
then $\bar \alpha = 2-\bar \gamma$.  Our best guess based on the extrapolations
to lower flux densities and lower frequencies is that $\bar \gamma =  -0.07$ and therefore
we use $\bar\alpha=2.07$ as our fiducial value.

Since clustering, according to \citet{dimatteo02}, 
is the dominant source of fluctuation power we have
$\xi = \sigma_\alpha/\Delta \alpha^2$.  For our fiducial model
we take $\Delta \alpha=\delta \gamma = 0.2$, consistent with \citet{cohen04}.   
The value for $\sigma_\alpha$ is less clear.  As \citet{zaldarriaga04}
argue, if the sources are all tracing the same dark matter distribution,
even with different biases, then they would be perfectly correlated
with each other; i.e., $\sigma_\alpha = \infty$.  Certainly on large scales
we expect this to be a good approximation.  In the context
of the halo model \citep{cooray02}, the relevant length scale is the size of a typical halo
hosting these sources.  How this translates to an angular scale depends on distance to
the source.  

We have not attempted to calculate $\sigma_\alpha$ here, but merely point out
that even taking the conservative assumption that $\sigma_\alpha = \Delta \alpha$
we get $\xi = 5$.  Because, as we will see, we can afford to be even more conservative,
we set our fiducial value of $\xi$ even smaller to $\xi= 1$.

We note that although for low $S_{\rm cut}$ clustering may be the dominant source of fluctuation power, 
Poisson fluctuations may remain as the dominant source of decorrelation, especially if $\sigma_\alpha >>
\Delta \alpha$.  In this case though, $\xi$ would still be greater than $1/\Delta \alpha$ and therefore
even more coherent than in our fiducial model.

We also note that although the extrapolations from current observations of sources in the relevant
wavelengths are over many decades in intensity, we are, of course, not stuck with these
extrapolations.  With the type of experiments we consider here, we will be able to study large
numbers of resolved sources with flux densities just above $S_{\rm cut}$.  

\subsection{Extragalactic free-free emission}

The low-frequency radio background produced by ionized electrons, via the free-free emission,
is known to be a significant foreground for 21 cm studies \citep{oh99,oh03,cooray04}.
The free-free emission coefficient can be written as
\begin{eqnarray}
\epsilon_\nu &=& 5.4 \times 10^{-39} n_e^2 T_e^{-1/2} g_{ff}(\nu,T_e)
e^{-h\nu/kT_e} \nonumber \\
&& \quad \quad \quad \, {\rm ergs \; cm^{-3}\; s^{-1}\; Hz^{-1}\;
  sr^{-1}} \, ,
\label{eqn:free}
\end{eqnarray}
where the Gaunt factor can be approximated in the radio regime as
$g_{ff}(\nu,T_e) \approx 11.96 T_e^{0.15} \nu^{-0.1}$ \citep{lang99}.  
The cumulative specific intensity is simply $I_\nu = \int
d\chi \epsilon_\nu(z) \bar{x}_e^2(z) \\ C(z)/(1+z)^4$, where $\chi$ is the
conformal time and $\nu(z) = \nu_{\rm obs}(1+z)$ with $\nu_{\rm obs}$ the observed frequency. 
The brightness temperature is $T_b = c^2 I_\nu/2\nu^2_{\rm obs}k_B$
and the electron fraction $\bar{x}_e(z)$ captures the mean
ionization history of the IGM. 
The clumping factor of the electron
density field is defined as $C(z) = \langle n_e^2\rangle/\langle n_e \rangle^2$. 
When estimating the free-free background and its spatial fluctuations, this is
the most uncertain parameter. Initial estimates of the free-free background suggest that gas clumping boosts
the specific intensity by a factor from order unity at $z\sim20$ to over 100 at $z \la 3$ \citep{oh99}.

A variety of models from the literature considered in \cite{cooray04} suggest that the free-free background
is uncertain by at least 2 orders of magnitude. Among these models, a favorite is to
estimate the clumping from the large-scale matter distribution by, for
example, assuming that ionized electrons reside in virialized dark
matter halos \citep{haiman01,benson01,gnedin97} though such models are only
expected to yield a minimal estimate for $C(z)$, because the models do not include recombinations
within galaxies (where most ionizing photons are probably absorbed) or
the biased distribution of ionizing sources.  Numerical simulations by \citet{gnedin00} 
show  that these two factors strongly affect the effective clumping
factor early in the reionization process, when ionizing photons are
confined to the dense regions around galaxies.
The second approach, taken by \citet{oh99}, is to construct a
model for the production rate of ionizing photons $\dot{n}_{\rm ion}$
and assume that, throughout the universe, ionization equilibrium is a
good approximation.  The quantity $\bar{x}_e^2 C(z)$ is then fixed through
the relation
\begin{equation}
\dot{n}_{\rm ion} = \bar{x}_e^2 (z)n_e^2(z) C(z) \alpha_{\rm rec},
\label{eq:nion}
\end{equation}
where $\alpha_{\rm rec}$ is the recombination coefficient.
This approach predicts stronger clumping than models based on the IGM gas as it
includes recombinations inside galaxies.  These dominate the clumping
so long as the escape fraction of ionizing photons is small.
However, this method depends on the assumed (and uncertain)
ionizing photon production rate. It also does not include collisional
ionization in massive halos at low redshift, though this is expected to be  small.

The results related to these two approaches and a comparison to results from numerical
estimates of gas clumping are summarized in \cite{cooray04}. 
In Fig.~\ref{fig:foreg}, we make use of the high-end estimate of free-free anisotropy spectrum based on a model related to
ionization equilibrium and using the star-formation history derived by \citet{hernquist03}.
In addition to the angular power spectrum, the foreground model requires information related to the coherence of fluctuations between
frequencies. Given that the extragalactic free-free emission is dominated by point sources at $z \sim 3$ which follow a biased
distribution with respect to the linear density field, the angular power spectrum is dominated by the clustering nature of this
point source distribution. The parameter of interest, $\xi$, then takes the value of $\sigma_\alpha/\Delta \alpha^2$ where $\alpha$ is
the frequency dependence determined by a combination of the Gaunt factor and e$^{-h \nu/k T_e}$. Even accounting for
variations in the electron temperature by an order of magnitude, from 10$^4$ to 10$^5$ K, the spectral index, 
as a function of frequency, varies slowly, since $h \nu \ll k T_e$. We take this small variations into account by 
setting $\xi=35$ (with $\Delta \alpha \sim 0.03$ assuming $\sigma_\alpha \approx \Delta \alpha$). This value is 
significantly high when compared to, say, radio point sources with a value of $\xi=1$ in our fiducial model. The difference is due to the
fact that observed spectral indices of point sources are rapidly varying from one source to another while free-free frequency
spectra, from source to source, is expected to be very uniform due to the nature of free-free emission.

While the anisotropy spectrum and frequency coherence are just estimates based on model calculations,
unfortunately, there are no observational measurements of the free-free background either in terms of mean
brightness temperature or spatial fluctuations. The planned Absolute Radiometer for Cosmology, Astrophysics, and
Diffuse Emission (ARCADE) experiment may be able to study some aspects of free-free background at frequencies
around a few GHz. Also, low-frequency experiments such as SKA will have a high frequency
imaging capabilities such that one can use data at frequencies just above 1420 MHz to determine the
extent to which free-free emission can be a contaminant at low radio frequencies.
While the free-free background can be a problem, in \citet{cooray04} it was shown that the
integrated signal is dominated by low-$z$ sources (at redshifts less than 3). Also, a large
fraction of the free-free fluctuations comes from relatively bright point
sources that can be cleaned from the maps.
Given that most of the emission is at low redshifts, when compared to 21 cm fluctuations from the reionization era,
the free-free spectrum is smooth in frequency. Thus, multi-frequency differencing is expected to
remove the smooth background radiation but not 21 cm radiation as the two fields are strongly uncorrelated given 
the disjoint in redshift distributions; we do expect some free-free emission from the era of reionization, but this was
shown to be substantially below the noise level of upcoming experiments in \citet{cooray04}. 

If the fraction of emission at $z \sim 10$ were  to be higher for some reason, 
then the confusion with free-free fluctuations may become substantial; for free-free
fluctuations to dominate 21 cm anisotropies, even with optimal multifrequency cleaning, we require the fractional
contribution to total emission from $z \sim 10$ to be at the same level as $z \sim 1$, which we do not consider to be physically
possible. Thus, it is unlikely that after multifrequency cleaning, free-free fluctuations will be the dominant residual contamination
of 21 cm maps.

\subsection{Galactic Backgrounds}
  
While radio point sources and ionized halos that emit free-free emission are expected to produce small-scale foreground anisotropy
structure at low radio frequencies, the large angular scale fluctuations are expected to be dominated by the synchrotron emission
within the Galaxy. The synchrotron background is moderately well understood, when compared to say low frequency faint
radio sources or the free-free background, given that it also can affect temperature anisotropy measurements of
the CMB at frequencies around 30 GHz. In addition to temperature anisotropies, the highly sensitive CMB polarization
observations, with substantially smaller signals, also require a thorough understanding of the synchrotron
background from the Galaxy.

In Fig.~\ref{fig:foreg}, we make use of a ``middle of the road'' estimate for the Galactic synchrotron fluctuations,
following \citet{tegmark00b}.  
The synchrotron spectrum has an angular power spectrum that scales with
multipole as $C_l \sim l^{-2.4}$ \citep{tegmark00b}, though this slope is rather
uncertain with observational measurements, e.g. from the 408 MHz Haslam map, 
varying from -2.5 to -3.0 down to angular scales of  a degree. 
In addition to the  spatial spectrum, we take a frequency spectrum that scales as $\nu^{-\alpha}$ and take $\alpha=2.8$
following the data from \citet{platania98}. The frequency spectrum, however, varies across the sky with $\Delta \alpha \sim$
0.15 \citep{tegmark00b}, though information related to $\sigma_\alpha$ is not yet available from the data. We take a conservative
estimate of 0.1 and set $\xi \sim \sigma_\alpha/\Delta \alpha^2$, as large scale clustering dominates the anisotropy spectrum,
to be 4 in the case of Galactic synchrotron fluctuations.

Note that the synchrotron amplitude is set based on foreground analysis of CMB experimental data at 19 GHz (cross-correlated with
low frequency maps)  where the power variance of synchrotron fluctuations was estimated to be $52 \pm 17$ $\mu$K 
at an angular scale of 3 degrees
by \citet{deoliveira-costa98}. Though fluctuation power decreases with increasing
angular resolution, the amplitude for these fluctuations is such that the synchrotron 
emission from the Galaxy
dominates temperature fluctuations in all angular scales of interest. 

Fortunately, the synchrotron emission can
also be contained in upcoming 21 cm observations given the fact that this power is expected to be
smooth over a wide band in frequency space while 21 cm fluctuations related to neutral Hydrogen at redshifts around
reionization will vary significantly.

In addition to synchrotron emission, the Galactic free-free background is also expected to confuse low-frequency
radio measurements of the neutral Hydrogen distribution around the era of reionization. Following \citet{tegmark00b}, we include this 
contribution as a separate component with a power spectrum that scales as
 $C_l \sim l^{-3.0}$ and a frequency spectrum with $\nu^{-2.15}$. As in the case of extragalactic free-free background,
we assume a large frequency coherence and set $\xi=35$ when $\Delta \alpha=0.03$. Note that the Galactic free-free background has a
similar spatial power spectrum as that of the synchrotron background, but
with a lower amplitude. While such a background can be ignored, given the lower amplitude compared to
Galactic synchrotron, we include this component due to the difference in frequency dependence and the fact that it has the
same frequency scaling as that of the extragalactic free-free component. By doing so, we can investigate how differences in frequency 
scaling and the power spectrum scaling  can complicate a potential frequency cleaning attempt.

\section{Experimental setup}
\label{exp_setup}

For an interferometer type experiment, the measured flux in a given visibility $V(\vecu,\nu)$,
can be expressed as 
\be
V(\vecu,\nu)={\partial B_{\nu}\over \partial T}\int d\bn T(\bn,\nu) A(\bn,\nu) e^{2\pi i\vecu\cdot\bn},
\ee
where $\vecu$ corresponds to a point in the $u-v$ plane, $T(\bn)$ is the sky temperature,
$A(\bn)$ is the primary beam and ${\partial B_{\nu}\over \partial T}$ converts temperature to flux (see e.g.
\citealt{white99}).
The visibilities will then consist of a convolution of the Fourier transform of $T(\bn)$
\be
\label{eq:visib}
V(\vecu,\nu)={\partial B_{\nu}\over \partial T}\int d^2u' \tilde T(\vecu',\nu)\tilde A(\vecu-\vecu',\nu)
\ee
with $\tilde T(\vecu)$ and $\tilde A(\vecu)$ the Fourier transforms of $T(\bn)$ and $A(\bn)$
respectively.
The typical width of $\tilde A(\vecu,\nu)$ will be set by the size of the dishes $d$, so that
$\Delta u=d/\lambda$ and visibilities closer than this in the $u-v$ plane will be
strongly correlated. 
The Fourier component $\tilde T(\vecu)$ is basically the flat-sky version of the
spherical harmonic transform of the sky temperature, $a_{l m}$ (for the correspondence see e.g.
\citealt{santos03}). Its power spectrum is given by
\be
\langle\tilde T(\vecu,\nu)\tilde T^*(\vecu',\nu')\rangle=\delta^D(\vecu-\vecu')C_{l=2\pi|\vecu|}(\nu,\nu').
\ee

Finally, we can relate the power in the visibilities with the usual angular power spectrum
(with the relation $l=2\pi|\vecu|$):
\bea
\label{eq:vispower}
& &\langle V(\vecu,\nu_1) V^*(\vecu,\nu_2)\rangle=
\left({\partial B_{\nu_1}\over \partial T}\right)
\left({\partial B_{\nu_2}\over \partial T}\right)\nn
&\times&\int d^2u' \tilde A(\vecu-\vecu',\nu_1)\tilde A^*(\vecu-\vecu',\nu_2) C_{l=2\pi|\vecu'|}(\nu_1,\nu_2)\nn
&\approx& C_{l=2\pi|\vecu|}(\nu_1,\nu_2)\left({\partial B_{\nu_1}\over \partial T}\right)
\left({\partial B_{\nu_2}\over \partial T}\right)\nn
&\times&\int d^2u' \tilde A(\vecu-\vecu',\nu_1)\tilde A^*(\vecu-\vecu',\nu_2),
\eea
where the last line is a reasonable approximation since the power spectrum $C_l$ is quite smooth.
From this expression we can see that, because the beam size depends on the frequency,
the visibilities will be less correlated across
frequency than the underlying signal on the sky \citep{dimatteo02}.
Basically the width of $\tilde A(\vecu)$ increases with frequency so that visibilities at higher
frequencies will have contributions from a larger range of $\tilde T(\vecu)$ (equation \ref{eq:visib}).
This is particularly
important in the foreground case,
since their correlation structure is crucial for the cleaning process. Using 
$\int d^2u\ \tilde A({\bf u},\nu)=1\sim (\Delta u)^2 \tilde A(0)$ so that
$\int d^2u^\prime \ |\tilde A({\bf u}-{\bf u}^\prime,\nu)|^2\sim 1/(\Delta u)^2$, the
foreground correlations can be written as
\be
{\langle V^{ii}(\vecu,\nu_1) V^{ii*}(\vecu,\nu_2)\rangle\over
\sqrt{\langle |V^{ii}(\vecu,\nu_1)|^2\rangle\langle |V^{ii}(\vecu,\nu_2)|^2\rangle}}\approx
I^{ii}_{l=2\pi|\vecu|}(\nu_1,\nu_2){\nu_1\over\nu_2},
\ee
where $I^{ii}_l(\nu_1,\nu_2)$ is given by equation (\ref{corr_def}) and $\nu_2>\nu_1$.
Even if $I^{ii}_l(\nu_1,\nu_2)=1$ the correlation can still be much smaller than 1, making foreground
removal extremely difficult. It will therefore be essential for this type of experiment
to generate an effective beam size that is constant across frequency. 
These considerations will place stringent constraints on experimental design and observing
strategy.  Only with sufficiently dense sampling and with the primary beam known sufficiently
accurately can higher frequency visibilities be deconvolved to match the lower frequency visibilities.
From now on
we will assume this is possible and drop the explicit frequency dependence of the beam.

\begin{figure}
\begin{center}
\plotone{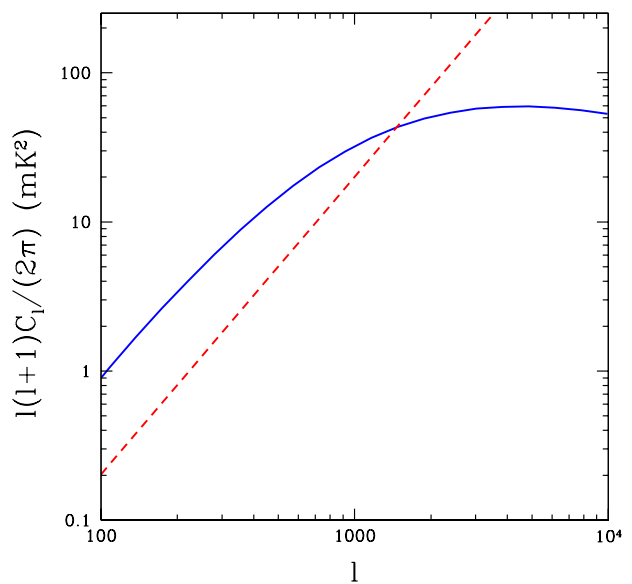}
\caption{The noise power spectrum assumed in our analysis (red dashed line).
Blue solid line - the 21cm signal at z=9.2 (140 MHz) with $\Delta\nu=1$ MHz.}
\label{fig:noise}
\end{center}
\end{figure}

The corresponding noise in each visibility is given by 
\citep{rohlfs04}
\be
\label{vis_n}
\langle |N({\bf u})|^2\rangle =   \left({2 k_B T_{sys} \over
	A_{dish}} \right)^2 \ {1\over \Delta {\nu} t_v}  
\ee
where  $T_{sys}$ is the system temperature, $\Delta {\nu}$ is the
bandwidth, and $A_{dish}$ is the area of each individual antenna in
the array.
From equation (\ref{eq:vispower}), the noise angular power spectrum will then be
\be
C_l^N = \left ( {\lambda^2 T_{sys} \over  A_{dish}} \right )^2 \  
{1\over \Delta {\nu} t_v \int d^2u^\prime \ |\tilde A({\bf u}-{\bf u}^\prime)|^2}\, .
\ee
Moreover, assuming the baselines are arranged so that the Fourier coverage is uniform, the
time spent for each visibility will be roughly constant. At a given instant, the 
experiment will observe an area in Fourier space given by $N_b(\Delta u)^2$, where $N_b$
is the number of baselines. The observation time per unit area will then be 
\be
{t_v\over N_b(\Delta u)^2}={t_0\over\pi u_{max}^2}, 
\ee
where $u_{max}$ corresponds to the largest
visibility observed and $t_0$ is the total time spent observing a given patch of the sky.
This way, using $N_b\sim N_{dish}^2/2$,
\be
C_l^N={T_{sys}^2 (2\pi)^3\over \Delta\nu t_0 f^2_{cover}l_{max}^2},
\ee
where $f_{cover}$ is the fraction of the total area covered by the dishes:
\be
f_{cover}\equiv{N_{dish} A_{dish}\over A_{total}}={N_{dish}(\Delta l)^2\over l_{max}^2}.
\ee
For our analysis, we used $f_{cover}=0.034$, $l_{max}=5000$, $T_{sys}=200$K and $\Delta\nu=1$MHz.
We choose the total observation time to be $t_{total}=4$ months.
To specify $t_0$ we took into account that it might be more efficient to observe different
patches of the sky in order to reduce sample variance. The optimal observation time for one
patch, given the above values, is $t_0\sim 1\ $month. Note that by optimizing the signal to noise
(signal/noise $\sim$ 1), the expected errors will scale as $\sim 1/f_{cover}/\sqrt{t_{total}}$, so that
it should be straightforward to extrapolate our results to other experiments (see \citealt{morales04}
for a discussion on the design of Epoch of Reionization observatories).
The correlation length across multipoles is set by the dish size, which we assumed to be
$d\sim 33\,$m corresponding in multipole space to $\Delta l=2\pi \Delta u\sim 100$ at $z=9$. Figure \ref{fig:noise}
shows the noise level we use, together with the expected signal at 140MHz.
  
\section{Error forecast}
\label{forecast}

Our goal is now to calculate the expected error in the 21cm power spectrum, taking into
account  the experimental noise and foreground contamination.
Given a set of measurements $\{y\}$, the information of how well we can measure a set of
parameters $\{p\}$, is contained in the Fisher information matrix \citep{kendall79}
\be
F_{ij}=\left\langle{\partial^2\log{\cal L}\over \partial p_i\partial p_j}\right\rangle,
\ee
where ${\cal L}(y,p)$ is the likelihood of parameters $\{p\}$ given observations $\{y\}$
and the matrix is to be evaluated at the true values of the parameters $\{p\}$.
For now, we consider these parameters to be the 21cm auto power spectra binned in multipole and
frequency space plus the foreground model parameters.
The Cram\'er-Rao inequality states that $(F^{-1})_{ij}$ is the smallest variance that any unbiased
estimator of the parameters can have (see \citealt{tegmark97c} for a review).

For an interferometer, we can take the set of measurements $\{y\}$ to be the measured 
visibilities, $V(\vecu,\nu)$.
Assuming all fluctuations are Gaussian with zero mean ($\langle y\rangle=0$) the
Likelihood is entirely specified by the covariance matrix 
${\bf C}\equiv \langle y y^t\rangle=\langle V(\vecu,\nu) V^*(\vecu',\nu')\rangle$.
Note that the brightness temperature is not
exactly Gaussian (see \citealt{furlanetto04b}). In order to obtain large values of the temperature 
we need regions with large densities, but these are typically ionized 
giving $T_{21}=0$, so that the Gaussian distribution will be cut for large values. However, 
as long as we are probing scales much larger than
the effective patch size, the signal will be an average over many bubbles and 
we expect the Gaussian assumption to be a good approximation
for the Fisher matrix analysis.
We also make the approximation that the foreground fluctuations are Gaussian distributed.
While this assumption holds true for high-redshift contributions, 
such as free-free point sources, foregrounds
such as synchrotron emission within the Galaxy are highly non-Gaussian over large areas on 
the sky. Our Gaussian assumption is then expected to
underestimate the effect of foregrounds at large angular scales corresponding to tens of degrees.
However, here we will consider measurements of 21 cm fluctuations at a degree scale and below 
so that the Gaussian approximation is expected to be a reasonable one.

The covariance matrix can be quite complex due to the cross-correlations between different $\vecu$.
To proceed further we assumed a bin size in $l=2\pi|\vecu|$ space such that this cross-correlation
is negligible. The matrix ${\bf C}$ will then become block diagonal, making the Fisher matrix
analysis more straightforward. As discussed in the previous section, it will be necessary to
create an effective beam such that the correlation length is constant across frequency.
We choose $\Delta l=100$ which, for the experimental setup described above (\S \ref{exp_setup})
makes the values almost uncorrelated among different $l$.
The number of independent measurements at a given $l=2\pi|\vecu|$ is then
$N_l={\pi l t_{total}\over\Delta l t_0}$.

We will proceed the analysis making reference to the angular power spectrum instead of the 
power in the visibilities,
knowing that the two can be related through equation (\ref{eq:vispower}).
The covariance at a given $l$ between two frequency bins $(i,j)$ is:
\be
{\bf C}^l_{i j}=C^f_l(\nu_i,\nu_j)+C^s_l(\nu_i,\nu_j)+
C_l^N(\nu_i,\nu_j)\delta_{ij}\, .
\ee
The corresponding Fisher matrix is then,
\be
{\cal F}_{a b}\approx\sum_l {{N_l\over 2}Tr\left[({\bf C}^l)^{-1}{\partial {\bf C}^l\over\partial p_a}
({\bf C}^l)^{-1}{\partial {\bf C}^l\over\partial p_b}\right]},
\ee
where $\{p_a,\,p_b\}$ are the parameters we wish to estimate.

\subsection{Power Spectrum}

In this section we investigate how well we will be able to measure the 21cm power spectrum.
To better understand the impact our foreground models will have on foreground removal, we
start by just considering one type of foreground (point sources) and assume that 
all its statistical properties are known.
Our parameters are then: 
$p_a=C_l^{21}(\nu_a,\nu_a)$ and we assume that the correlation of this
signal across frequency, e. g. ${C_l^{s}(\nu_a,\nu_b)/\sqrt{C_l^{s}(\nu_a,\nu_a)
C_l^{s}(\nu_b,\nu_b)}}$, is known from theory.

We can calculate an approximate expression for the expected error in $C_l^{21}(\nu_i,\nu_i)$ which
will make it easier to understand the results from the Fisher matrix analysis.
If the foregrounds were perfectly correlated across frequencies 
($a^f_{l m}(\nu_2)=\gamma a^f_{l m}(\nu_1)$) and the signal uncorrelated, then
clearly by subtracting the two maps ($a_{l m}(\nu_2)-\gamma a_{l m}(\nu_1)$) the
foregrounds would be completely removed.
A decrease in the foreground correlation translates into fluctuations in
$\gamma$ so that the foreground subtraction is not as effective.
Using only two frequency bins, the corresponding error of the 21cm power spectrum is then 
(see \citealt{zaldarriaga04})
\be
\label{var21}
(\Delta C^{21}_l)^2\approx {2\over 2l+1}\left[C^{s}_l+C^N_l+
C^f_l(1-I)\right]^2,
\ee
where $I$ is the correlation coefficient and we have assumed the signal power spectrum
to be the same in both frequencies.  Cleaning in this case is very effective, with
foreground contamination down by $C^f_l(1-I)$.

We want to reconstruct the signal power spectrum as a function of frequency and so now drop this
assumption of frequency independence.  Without this assumption, two channels are
now only sufficient for constraining a linear combination of the signal power spectra, not the
signal power spectra themselves.  However, with $N$ channels, $N(N-1)/2$ different foreground-cleaned 
maps can be made (one for each pair of maps) which is more than the $N$ signal
power spectra to be determined if $N > 2$, making the reconstruction possible.  The
more channels, the more pairs of maps and the more estimates of the power spectra we have.  
If all of these map pairs had similar, and independent, information about the signal power
spectra, then we would expect the errors on each power spectrum to improve as $\sim \sqrt{(N-1)/2}$.
Actually, the errors are correlated, and the difference maps made from larger frequency
separations have greater residual foreground contamination, so the improvement is slower
than $\sqrt{(N-1)/2}$ if one increases $N$ by increasing the frequency coverage.

The growing decoherence of the foregrounds over large frequencies
limits how well an expanded frequency coverage can improve the
reconstruction.  Likewise, the coherence of the {\em signal} over
very small frequency separations limits how well a refined binning
can improve the reconstruction.  With a fine enough binning, moving
to finer binning no longer increases the effective number of independent
frequency bins.  Figure 7 shows the expected foreground and signal
correlations across frequency.  Although, as expected, the foreground
is much more strongly correlated than the signal, there is a
non-negligible signal correlation.  The moderate correlation strength
for frequency separations equal to the bandwidth (about 30\%) suggests 
that there will be some improvement in reconstruction from making
bands finer than 1 MHz, but that 1 MHz is near the point of highly
diminished returns, at least for $l \la 1000$.
Binning an order of magnitude more finely than our
nominal 1 MHz bins would also result in the added challenge of having
to take into account redshift distortions due to the peculiar
velocities of the neutral gas. With 1 MHz binning the redshift
distortions are negligible.
\begin{figure}
\begin{center}
\plotone{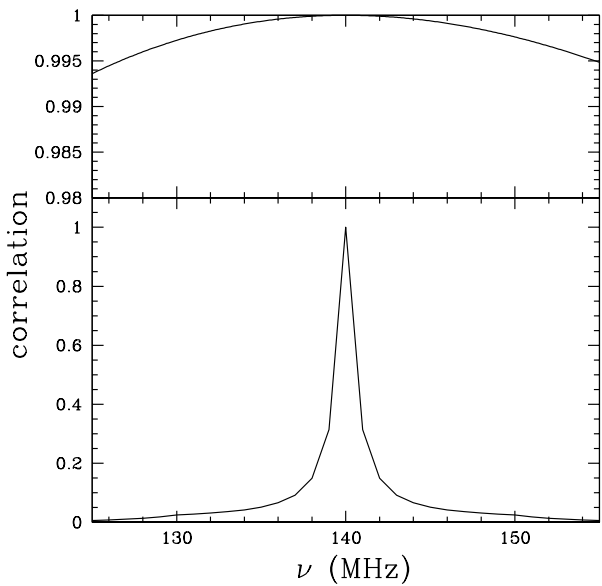}
\caption{Expected frequency cross-correlations for
foregrounds (top) and signal (bottom) at $l=1000$ ($\Delta\nu=1$ MHz).}
\label{fig:correlat}
\end{center}
\end{figure}

\begin{figure}
\begin{center}
\plotone{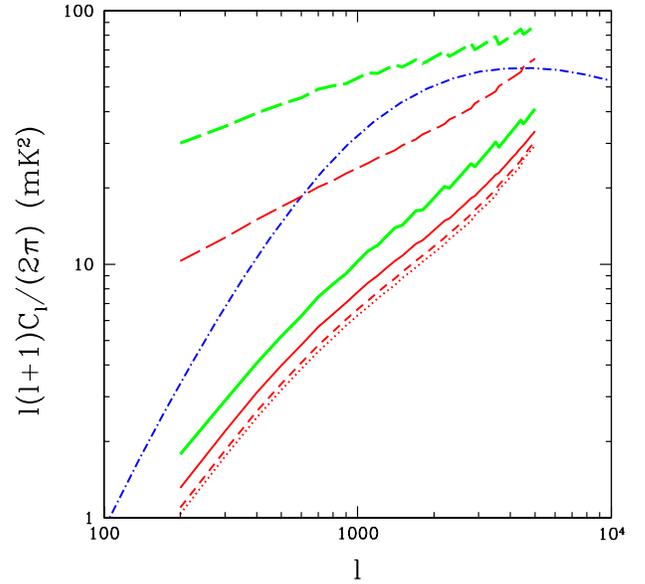}
\caption{Expected error on the power spectrum at 140MHz (1MHz bin size), assuming only point
sources contamination (no free parameters).
Blued dot-dashed line - the 21cm signal. All other curves show the recovered errors for different
assumptions about the point source model, using 40 frequency bins (thin red lines) and
10 bins (thick green lines).
Dotted line - error with no foregrounds.
Solid curve - assuming a Gaussian correlation across frequency.
Short-dashed curve - point sources fully correlated. Long-dashed curve - assuming a constant
correlation. This constant correlation case uses $I\sim 1-3\times 10^{-5}$ (the value of
the Gaussian at 1MHz separations).}
\label{fig:error1f}
\end{center}
\end{figure}

In Figure \ref{fig:error1f} we show the expected error in the 21 cm power spectrum
taking into account the experimental setup described above and only assuming point
source contamination. We used the point source parameters from Table~\ref{tab:fiducial}.
The red thin lines use information from 40 frequency bins (from 120MHz
to 160MHz) while the green thick curves only use ten bins (from 135MHz to 145MHz). 
The larger the number of bins, the smaller is the recovered error, as expected from
the above discussion.
Since foreground subtraction is difficult at frequency
separations less than 1MHz, the best way to improve on this result will be to expand 
the frequency range, although this can be problematic due to the man-made electromagnetic
interference. Note however that the recovered error using our fiducial point source model (thin solid line) is
already quite close to the ideal situation with no foreground contamination (dotted line).
Assuming perfectly correlated foregrounds (short-dashed line), the errors come very close to the no foreground
case; they would be identical were it not for the signal correlation.
Comparing the solid and long-dashed curves, we also see
that the Gaussian form of the foreground correlations (e.g.
equation \ref{gauss_corr}) allows the foreground removal to be more efficient than in
the constant correlation case. This is an important result that we will explore in the next section.

\begin{figure}
\begin{center}
\plotone{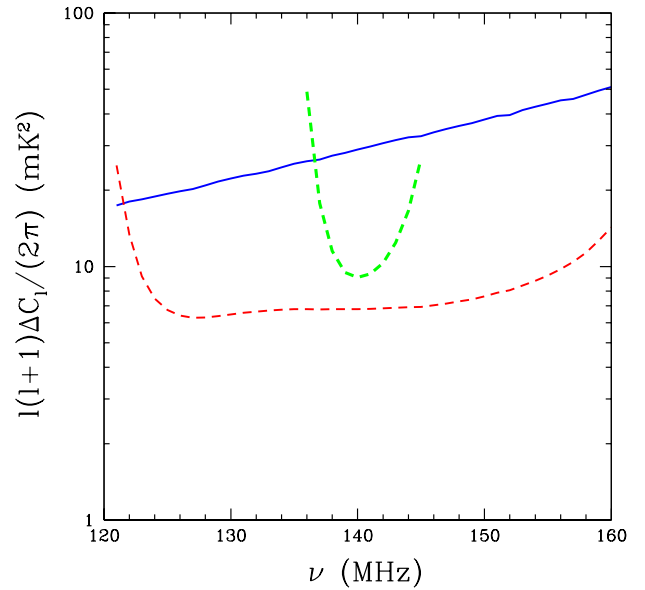}
\caption{The 21cm signal at $l=1000$ as a function of frequency (blue solid curve) and the corresponding errors. We are
assuming only point source contamination (parameters from Table~\ref{tab:fiducial}).
Red thin dashed line shows the recovered errors using 40 frequency bins. Green thick dashed line shows the
same for 10 bins.}
\label{fig:errorfreq}
\end{center}
\end{figure}

Note also that we are presenting this result
for a bin at the middle of the frequency range. 
Figure \ref{fig:errorfreq} shows the power spectrum and its errors at $l=1000$,
versus the frequency, again assuming only point source contamination.
As we move closer to the end of the frequency interval,
the error increases, since the number of neighboring channels decreases (also, for lower
frequencies, the foreground contribution is typically larger). However, using the full
frequency range still provides great statistical power when distinguishing between
different reionization models as we will see later in \S \ref{reion}. 

\subsection{Dependence on assumptions about frequency coherence and its
functional form}

\begin{figure}
\begin{center}
\plotone{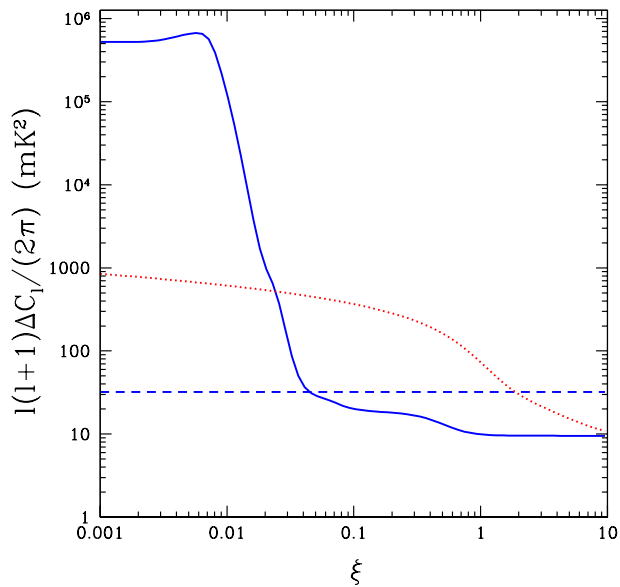}
\caption{Dependence of the expected error on the foreground correlation length (at z=9.2 and $l=1000$) using
a Gaussian correlation function (blue solid line). Red dotted line shows the same for the thick solid curve
in Figure \ref{fig:fct} (e.g. the worst case scenario).
Blue dashed line shows the signal.}
\label{fig:err_xi}
\end{center}
\end{figure}

In general, the more coherent a foreground is, the easier it will be to remove it.
This can be seen in Figure \ref{fig:err_xi} where we plot the recovered error for
the power spectrum at $l=1000$ as a function of the correlation length, $\xi$ (solid curve).
As above, we are using a Gaussian form for the correlation function.
As expected, the situation gets worse as we move from the ideal case of perfect
foreground coherence (large $\xi$) to completely incoherent foregrounds (small $\xi$).
In the latter case, the foregrounds simply act as a huge noise component.
Foreground removal is shown to be effective as long as $\xi\gtrsim 0.1$ which,
if the point sources have a Poisson distribution, would translate into
$\Delta\alpha \lesssim 10$. If the point source fluctuation power is mainly
from correlations and we conservatively set $\sigma_\alpha = \Delta \alpha$,
then $\xi\gtrsim 0.1$ once again translates into $\Delta\alpha \lesssim 10$.  We see that
our extrapolations based on data with brighter point sources would have to be very far
off in order for the uncertainty in the 21cm fluctuation power spectrum to be
significantly increased.

The shape of the correlation function (equation \ref{corr_def})
also has a very important impact on how well foregrounds can be removed. In fact we have already seen in
Figure \ref{fig:error1f} that the (unphysical) assumption of constant cross-frequency correlations
results in larger errors, even though the foreground correlations between different frequencies
are always larger than in the Gaussian case (a similar effect, albeit in a different scenario, was
already noticed in \citealt{tegmark00b}).
Since the coherence function, $I(\nu_1,\nu_2)$, has not yet had its shape accurately measured, 
we decided to repeat the Fisher analysis for a variety of such functions. We can rewrite the
coherence function (equation \ref{gauss_corr}) as $I(\nu_1,\nu_2)=f\left[\log(\nu_1/\nu_2)/\xi\right]$,
where, for our fiducial model, $f(x)=e^{-x^2/2}$.
Although this function should behave like a parabola near the origin (
equation \ref{fct_corr}), with $f(0)=1$, $f'(0)=0$ and $f''(0)=-1$ \citep{tegmark98}, it does not
have to necessarily be a Gaussian, so that we also considered functions of the form
\be
f(x)=\left(1+{x^2\over 2 n}\right)^{-n}.
\ee
We recover the Gaussian case for $n=\infty$ (actually, for the typical range of values in $x$ we are
considering, $n=10$ is already a good approximation to a Gaussian). 

\begin{figure}
\begin{center}
\plotone{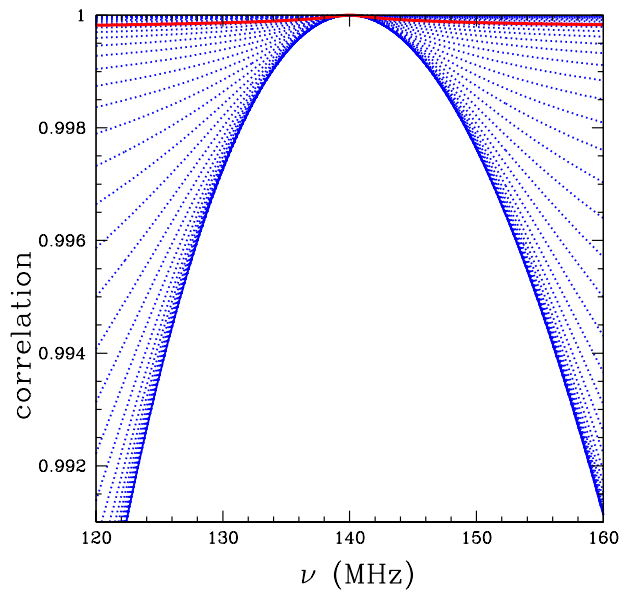}
\caption{Several functions for the foreground correlations of the type: $(1+x^2/2/n)^{-n}$ with
$x=log(\nu_1/\nu_2)/\xi$. Gaussian: n$\rightarrow\infty$ ($e^{-x^2/2}$). 
Red thick solid line gives the largest error and
is very similar to the constant case with a correlation at 1MHz separations identical to the Gaussian case.}
\label{fig:fct}
\end{center}
\end{figure}

\begin{figure}
\begin{center}
\plotone{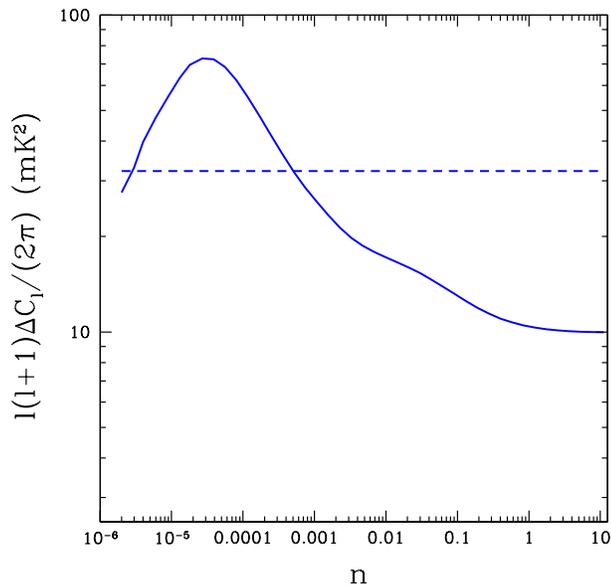}
\caption{The recovered error at $l=1000$ and $\nu=140$ MHz, assuming point sources contamination
and using the correlation functions from figure \ref{fig:fct} (blue solid line).
Dashed line shows the signal.}
\label{fig:fct_error}
\end{center}
\end{figure}

Figure \ref{fig:fct} shows the foreground correlations
for several values of $n$ (decreasing $n$ increases the values at the tails of the distribution).
In Figure \ref{fig:fct_error} we show the recovered error in the power spectrum at $l=1000$ using the
above functions. In this case we are only using 10 bins of 1 MHz each, centered around 140MHz (solid line). 
Notice that, although all the functions have larger cross-frequency correlations
than the Gaussian case, foreground removal is always worse than with the Gaussian form (large $n$).
The behavior of $f(x)$ away from the origin is quite important. As $n$ decreases, the second
derivative of $f(x)$ for $x\neq 0$ also decreases, so that the function approaches the linear
(and almost constant) case. This translates into an increase in the recovered error which peaks
at $n\sim 3\times 10^{-5}$. At this point, the correlation for separations of 1MHz (the bin size we
have been using) is still approximately the same as in the Gaussian case. As $n$ decreases further,
the frequency correlations increase (even at 1MHz separations) and the function converges to 1 everywhere
(foregrounds fully correlated), so that the recovered error decreases again.
Clearly, for foreground removal to be effective in the worst case scenario function, we need the correlation
length to be larger. Figure \ref{fig:err_xi} also shows the recovered error as a function of correlation
length for this worst case value of $n$ (red dotted lines).
In order to measure the signal we need $\xi>1$ which implies $\Delta\alpha < 1$.

\subsection{Simultaneous estimation of foregrounds and the 21cm power spectrum}

\begin{figure}
\begin{center}
\plotone{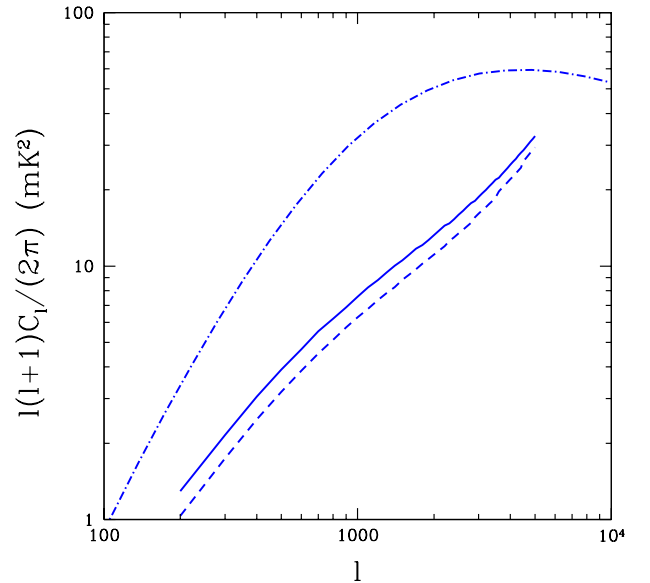}
\caption{Expected error on the power spectrum at 140MHz, when including point sources, 
Galactic synchrotron, extragalactic free-free and Galactic free-free (blue solid line). 
Dashed line gives the case with no foregrounds.
Dot-dashed line shows the signal.}
\label{fig:full_err}
\end{center}
\end{figure}

Up to now we have assumed that all the statistical properties of the foregrounds
are perfectly known. In fact, we hope to learn a lot about the spatial and frequency
dependence of foregrounds from experiments set to measure the 21cm signal.
Information about the spectral index of point sources at the frequencies of interest, 
for instance, can be obtained by using the large baselines available with the
interferometer experiments. This combined with observational data on
foregrounds at frequencies slightly higher than 1421 MHz 
should provide us with an improved foreground model
prior to the data analysis. However, because of the huge foreground amplitude compared to 
the 21cm signal, even very small
variations to the assumed model might be confused with the signal and bias the estimation. 
In particular, as we have seen
above, a good knowledge of the foreground frequency coherence structure in the range
of interest will be crucial to the analysis. Therefore, we will now allow for free
parameters in our foreground model and consider the case in which these parameters,
together with the 21cm power spectra are simultaneously estimated from the data.
Our foreground model is as given in equations ~\ref{eqn:fg1} with
\be
I_l(\nu_1,\nu_2)=\left(1+{\log^2{(\nu_1/\nu_2)}\over 2n\xi^2}\right)^{-n}.
\ee
The parameters to estimate are $p=\{A,\beta,\bar\alpha,\xi,\log(n)\}$ for each of the four foregrounds we are
considering (point sources, Galactic synchrotron, extragalactic free-free and Galactic free-free emission) and 
$p_a=C_l^{21}(\nu_a,\nu_a)$. 
We use 49 bins in $l$ space at each frequency (from $l=200$ to 5000 with 
$\Delta l=100$) and 40 bins of 1MHz from 120MHz to 160MHz ($z=7.9$ to 
$z=10.8$) giving a total of 1980 parameters (20 parameters for the 
foregrounds plus 1960 for $C_l^{21}(\nu_a,\nu_a)$). Each bin in $l$ space 
has $N_l={\pi l t_{total}\over\Delta l t_0}$ independent $a_{l m}$ 
measurements so that there will be a total of 642605 input data points in 
the form of $a_{l m}$ values.
We show in Figure \ref{fig:full_err} the estimated error on the power spectrum 
for the frequency bin of 140MHz from our full calculation.
Using these 40 frequency bins, the foreground removal is quite efficient with the errors 
only showing a small foreground contamination. 
Note however that these errors will increase
as we move toward the limits of the frequency range, as shown earlier in 
Figure \ref{fig:errorfreq}.

\section{Constraining the reionization model}
\label{reion}

\begin{figure}
\begin{center}
\plotone{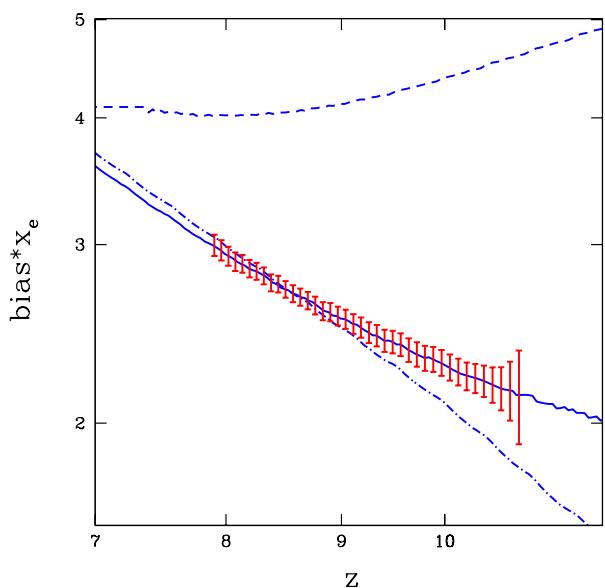}
\caption{Expected error on the product of the bias times the ionization fraction.
Models are the same as in Figures \ref{fig:xe} and \ref{fig:bias}.}
\label{fig:bx_err}
\end{center}
\end{figure}

Given a model of the perturbations in the ionization fraction we can ask how well
one can constrain the corresponding free parameters.
Typically, the ionization fraction power spectrum can be divided into two regions.
First, on large enough scales, we expect the fraction of ionized material to follow
the number of halos above a certain mass. The power spectrum will then be proportional
to the matter density, with a bias factor that is expected to be larger than one.
Second, as the scales approach the ionized bubble size, the power should be smoothed out
since there are no fluctuations from inside the bubbles (this small scale cutoff might be
quite extended if there is a wide distribution of bubble sizes). The transition from one
region to the other is model dependent. We are probing redshifts from
$z\sim 8$ to $z\sim 11$ up to $l\sim 1000$ corresponding to scales $x\gtrsim 9 {\rm Mpc}$
which we expect to be larger than the typical bubble size in our model.
We therefore model the ionization fraction power spectrum as
\be
P_{\delta_x \delta_x}(k,z)\sim b^2(z)P_{\delta \delta}(k,z)
\ee
and choose our free parameter to be $b_x \equiv \bar{x}_e b(z)$.
For large bias ($b(z)$), the main contribution to equation \ref{p21} is
then $b_x^2(z) P_{\delta \delta}(k,z)$. We have therefore a total of 60 parameters
in our model (one $b_x$ for each of the 40 frequency bins plus 20 foreground parameters).
Figure \ref{fig:bx_err} shows the expected error
from our full calculation. For comparison we also plot the expected $b_x$ from two
other reionization models with the same optical depth (see Figures \ref{fig:xe} and \ref{fig:bias}).
The recovered error is only at a few percent level so that, by measuring this parameter,
it should be possible to distinguish between different reionization histories.

\section{Summary}

The measurement of the 21cm radiation from neutral hydrogen at high redshifts will
provide unprecedent information about the epoch of reionization. Unfortunately this
signal has a huge foreground contamination, several orders of magnitude above the 21cm
fluctuations \citep{oh03}, making its detection a real challenge. In this paper we have presented a
detailed study of the foregrounds and the manner in which they will affect the measurement
of the 21cm perturbations.
We consider four types of foregrounds: diffuse Galactic synchrotron emission at low radio
frequencies, Galactic free-free, extragalactic free-free emission and extragalactic point
sources. We analyzed both the case when the foreground power spectra are known and the case when
their parameters must be determined from the 21cm data itself.

Our results are in general encouraging and show that we can achieve a high level of foreground
cleaning even when using quite conservative choices for foreground models both in terms of
foreground fluctuation amplitude and frequency coherence. The removal is aided by the
large frequency cross-correlation for foreground signals, although their
frequency dependence and spatial power spectra are also useful in distinguishing them from the
21 cm signal. The 21 cm signal correlations, between channels, also play an important role and basically 
set a minimum bandwidth to the frequency cleaning process in which foreground removal is still effective. 
We do not expect a binning finer than 1MHz to greatly improve the reconstruction.
Also, as expected, foreground removal improves as we increase the number of frequency bins with fixed bin size.
Using the full information from 40 bins of 1MHz each, it is possible to clean out the foreground signal almost
completely from the background 21 cm signal.

The most crucial feature in the cleaning process is the foreground correlation structure.
Concentrating on the point source contamination (which should be the most damaging
foreground due to its smaller coherence), we see that the Gaussian correlation
function provides the best results. In this case, limits on the spectral index
variation are well above expected values from point source models. Even
in the worst case scenario, when the function is nearly linear in
$\log\left(\nu_1/\nu_2\right)$, foreground removal is still effective as long as the spectral
index variations $\Delta\alpha\lesssim 1$, which is consistent with observations of point
sources \citep{cohen04}.
Our full statistical analysis allowed for variations in this
correlation structure as well as the spectral index and shape of the angular power
spectrum. Using a total of 40 bins from 120MHz to 160MHz and taking into account all
the foregrounds, we calculated the expected errors for a total of 1975 parameters. The results
show that foreground contamination will be minimal, allowing for a clear measurement
of 21cm power spectra.

We also investigated how well one can constrain the reionization model, by making full use 
of the three dimensional nature of the data (both in frequency and angular dependence). 
Although a detailed parameterization of the reionization model incorporating all 
possible variations to the reionization history is difficult, we expect certain 
features to be robust. On scales larger than the patch size, for instance, 
the 21cm power spectra should follow the shape of the projected dark matter
power spectrum.
Our analysis has shown that the large-scale bias of the neutral Hydrogen gas 
distribution, with respect to the density field, can be determined with a high precision. 
This allows for an accurate distinction between certain reionization models, even if 
they have the same optical depth.

The technical requirements to measure the 21cm signal, although challenging, should be 
within reach by proposed experiments such as SKA, LOFAR and PAST.
Instruments such as PAST can provide additional information related to foregrounds that can be
used to optimize SKA/LOFAR surveys. The big question addressed in this paper was to quantitatively
establish if foregrounds, when taken as a whole, would contaminate 21 cm measurements beyond repair. 
Our results show that the future for these experiments is quite promising: even fully taking into 
account the expected foregrounds, the 21 cm fluctuations can be measured well 
enough to be a powerful probe of reionization.

\acknowledgments
We thank S. Ananthakrishnan, Steven Furlanetto, Zoltan Haiman, Gil Holder, 
Miguel Morales, Rosalba Perna,
Max Tegmark and Xiaomin Wang for useful conversations.  
LK and MGS were supported by the  National Science Foundation under 
Grant No. 0307961 and NASA under grant No. NAG5-11098. 
MGS was also supported by the FCT-Portugal under grant
BPD/17068/2004/Y6F6.
AC was supported by the
Sherman Fairchild foundation and NASA NAG5-11985.

\bibliography{../biblio.bib}

\end{document}